\documentclass[useAMS,usenatbib,usegraphicx]{mn2e}
\DeclareMathVersion{bold}

\newcommand{\wjjj}[6]
{{
\left( 
\begin{array}{lcr} #1 & #2 & #3 \\#4 & #5 & #6 \end{array}
\right) 
}}

\title[Low CMB variance in WMAP]{A low CMB variance in the WMAP data} 
\author[Monteser\'\i n et al.]{C. Monteser\'\i n$^{1,2}$,
R.B. Barreiro$^1$, P. Vielva$^1$, E. Mart{\'\i}nez-Gonz{\'a}lez$^1$,
\and M.P. Hobson$^3$ and A.N.Lasenby$^3$   \\ 
$^1$ Instituto de F\'\i sica de Cantabria, CSIC-Univ. de Cantabria,
Avda. de los Castros s/n, 39005 Santander, Spain\\
$^2$ Dpto. de F\'\i sica Moderna, Univ. de Cantabria, Avda. de
los Castros s/n, 39005 Santander, Spain \\
$^3$ Astrophysics Group, Cavendish Laboratory, J.J. Thomson Avenue, Cambridge CB3 0EH  \\
}

\begin{document}
\maketitle

\begin{abstract}

We have estimated the CMB variance from the three-year WMAP data,
finding a value which is significantly lower than the one expected
from Gaussian simulations using the WMAP best-fit cosmological model,
at a significance level of 98.7 per cent.  This result is even more
prominent if we consider only the north ecliptic hemisphere (99.8 per
cent). Different analyses have been performed in order to identify a
possible origin for this anomaly. In particular we have studied the
behaviour of single radiometers and single year data as well as the
effect of residual foregrounds and 1/f noise, finding that none of
these possibilities can explain the low value of the variance.  We
have also tested the effect of varying the cosmological parameters,
finding that the estimated CMB variance tends to favour higher values
of $n_s$ than the one of the WMAP best-fit model. In addition, we have
also tested the consistency between the estimated CMB variance and the
actual measured CMB power spectrum of the WMAP data, finding a strong
discrepancy. A possible interpretation of this result could be a
deviation from Gaussianity and/or isotropy of the CMB.

\end{abstract}


\section{Introduction}
\label{sec:intro}

The study of the anisotropies of the cosmic microwave background (CMB)
constitutes one of the most powerful tools of cosmology, providing us
with very valuable information about the origin and evolution of the
universe. In particular, the determination of the CMB power spectrum
allows one to put tight constraints on the cosmological parameters. In
addition, the study of the CMB temperature distribution provides us
with a powerful test of the standard inflationary theory, since this
predicts that the CMB fluctuations should follow a homogeneous and
isotropic Gaussian distribution whereas alternative theories -- such
as non-standard inflation \citep{bar04} or topological defects
\citep{dur99} -- give rise to non-Gaussian fluctuations.

Given the interest of the subject, a large number of techniques have
been proposed for studying the Gaussianity and isotropy of the CMB
including, among others, the Minkowski functionals
(\citealt{col88,got90}), the bispectrum
(\citealt*{fer98,hea98,mag00}), properties of hot and cold spots
(\citealt{col87,mar00}), geometrical estimators
(\citealt*{bar01,dor03,mon05,mon06}), extrema correlation function
(\citealt{nas95,bar98,hea99}), wavelet analysis
\citep*{hob99,bar00,barh01,agh01}, bipolar power spectrum
\citep{haj03}, phase analysis (\citealt*{chi04}) and goodness of fit
tests \citep{ali05,rub06,cur07}.

Many non-Gaussianity analyses have been performed using the best CMB
data available up to date, which have been provided by the NASA
Wilkinson Microwave Anisotropy Probe (WMAP) satellite
(\citealt{benn03a}, \citealt{spe07}). The WMAP team carried out a
study of the CMB temperature distribution of the first year data
\citep{kom03} finding that the data were consistent with
Gaussianity. However, in subsequent works, a number of unexpected
results regarding the Gaussianity and/or isotropy of the CMB were
reported, including anomalies related to low multipoles
\citep{oli04,cop04,lan05a,sch04}, north/south asymmetries
\citep*{eri04,han04a,han04b,eri05,lan05b,ber06}, a cold spot
in the southern hemisphere \citep{vie04,muk04,cru05,cru06,mce05,cay05},
structure alignment \citep{wia06}, phase correlations
\citep*{chi03,col04} and to the amplitude of hot and cold spots
\citep{lar04}.

After the release of the three-year WMAP data in March 2006, similar
analyses have been carried out confirming the presence of the
anomalies in the data
\citep{mar06,mce06,vie07,cru07,cop07,ber07,eri07,wia08}.

In this work, we report a new anomaly in the WMAP data: a
significantly low value of the CMB variance with respect to the one
expected for the WMAP best-fit model. The outline of the paper is as
follows. In \S\ref{sec:tests} several quantities related to the
one-point density function (1-pdf) are studied for the WMAP data,
including the dispersion, skewness, kurtosis and Kolmogorov-Smirnov
distance. In \S\ref{sec:sigma0} the CMB dispersion from the WMAP data
is estimated, finding that its value is significantly low. The effect
on our results of possible residual foregrounds or systematics are
studied in \S\ref{sec:4}. In \S\ref{sec:explanations} possible
explanations for the anomaly are discussed, including the modification
of the cosmological parameters and a deviation of the CMB from
Gaussianity and/or isotropy. Finally, our conclusions are presented
in \S\ref{sec:conclusions}.
           
\section{Tests of Gaussianity based on the 1-pdf}
\label{sec:tests}

The NASA WMAP satellite was launched in the summer of 2001. The
first-year and three-years results were presented in February 2003 and
March 2006 respectively. WMAP observes at five frequency bands: K
(22.8 GHz, one receiver), Ka (33.0 GHz, one receiver), Q 
(40.7 GHz, two receivers), V (60.8 GHz, two receivers) and W (93.5
GHz, four receivers). All the data and products generated by WMAP can
be found at the Legacy Archive for Microwave Background Data Analysis
(LAMBDA) web site.\footnote{http://cmbdata.gsfc.nasa.gov}

Following the WMAP team (\citealt{kom03}) we have performed our
non-Gaussianity analyses on a noise weighted average of the Q, V and W
receivers, which provides a CMB map where the signal-to-noise ratio
has been increased. The K and Ka receivers are not included in the
combined map since they are largely contaminated by Galactic
foregrounds. In addition, the WMAP team has reduced the Galactic
contamination present in the Q, V and W map by performing a foreground
template fit as described in \cite{hin07}.
In particular, the combined Q+V+W map is constructed using these clean
data as
(\citealt{benn03b}):
\begin{equation}
\label{eq:combination}
T_c({\mathbf{x}}) = \sum_{j = 3}^{10} 
{T_j}({\mathbf{x}})~{w_j}({\mathbf{x}}),
\end{equation}
where $\mathbf{x}$ gives the position in the sky and the index $j$
correspond to the different receivers of the Q, V and W bands
(i.e., the indices 3 to 10 refer to the Q1, Q2, V1, V2, W1, W2, W3 and
W4 radiometers respectively).
The noise weight ${w_j}({\mathbf{x}})$ is defined as:
\begin{eqnarray}
\label{eq:noise}
w_j({\mathbf{x}}) = \frac{\bar{w}_j({\mathbf{x}})}{\sum_{j =
3}^{10}{\bar{w}_j}({\mathbf{x}})},~~ &
\bar{w}_j({\mathbf{x}}) = \frac{{N_j}({\mathbf{x}})}{{{\sigma_0}_j}^{2}}
\end{eqnarray}
where ${{\sigma_0}_j}$ is the noise dispersion per observation for
each radiometer given by \citealt{jar07}
and ${N_j}({\mathbf{x}})$ is the number of observations made by the
receiver $j$ at the position in the sky ${\mathbf{x}}$.

Although the data are provided at a
HEALPix\footnote{http://healpix.jpl.nasa.gov} \citep{gor05} resolution
of $n_{side}=512$, we degrade the data down to $n_{side}=256$ since the
smallest scales are dominated by noise. In addition, in order to avoid
the strong contamination present at the Galactic plane and the
emission coming from extragalactic point sources, only the data
outside the WMAP Kp0 mask \citep{hin07}, which corresponds
approximately to 76 per cent of the sky, has been used.  Finally, the
monopole and dipole outside the Kp0 mask have been removed.

In order to apply our Gaussianity test, we construct the normalised
temperature $u \left(\mathbf{x} \right)$, which is obtained by
dividing the data at each pixel by its corresponding expected
dispersion. Since the contribution from residual foregrounds in the
clean WMAP combined map outside the Kp0 is expected to be very small,
the main contribution to the data dispersion comes from the CMB signal
and the instrumental noise. The WMAP noise is very well approximated
by Gaussian white noise at each pixel characterised by a dispersion
$\sigma_n(\mathbf{x})$. Therefore, the normalised temperature is given
by
\begin{equation}
u \left(\mathbf{x} \right) = \frac{T_c \left(
\mathbf{x}\right)}{\sqrt{\sigma_{0}^{2}+\sigma_{n}^{2}\left(\mathbf{x}
\right)}}
\label{eq:norm_temp}
\end{equation}
where $\sigma_0$ is the CMB dispersion\footnote{Since the CMB is an
homogeneous and isotropic field, $\sigma_0$ is constant over the
sky. However, strictly speaking, there is a small dependence of
$\sigma_0$ with the position in the WMAP combined map due to the
different beams and noise weights used in the
combination. Nonetheless, these differences are very small and one can
assume, to a very good approximation, that the signal dispersion is
constant over the sky.}  and has been estimated as
\begin{equation}
\sigma_{0} = \sqrt{\sum_{l=2}^{\ell_{max}} {\frac{2\ell +1}{4 \pi}
C^c_{\ell}}} 
\label{eq:cmb_disp}
\end{equation}
The $C^c_{\ell}$'s in the previous equation correspond to the power
spectrum of the combined map assuming the best-fit model\footnote{ The
CMB simulations have been performed using the best-fit model -- for
the case of a $\Lambda$CDM model using only WMAP data --  given in the
first version of the papers of the three-year WMAP data realease
(\citealt{spe07}, see table 5 of arXiv:astro-ph/0603449v1). Although
the values of the cosmological parameters have been slightly modified
in the final version of the WMAP papers, we have tested that these
small changes do not affect the results of this work.}  to the WMAP
data and taking into account the different beams and noise weights
used in the combination (see Appendix \ref{ap:cl_comb_teo}). The
maximum $\ell$ that we have considered is $\ell_{max}=2.5 n_{side}$.

If the data follow a Gaussian distribution -- and our assumptions about
the underlying cosmological model and the instrumental noise are correct
-- the normalised temperature $u(\mathbf{x})$ should follow a Gaussian
distribution with zero mean and unit dispersion N(0,1). 

\begin{figure}
\begin{center}
\includegraphics[angle=0, width=\hsize]{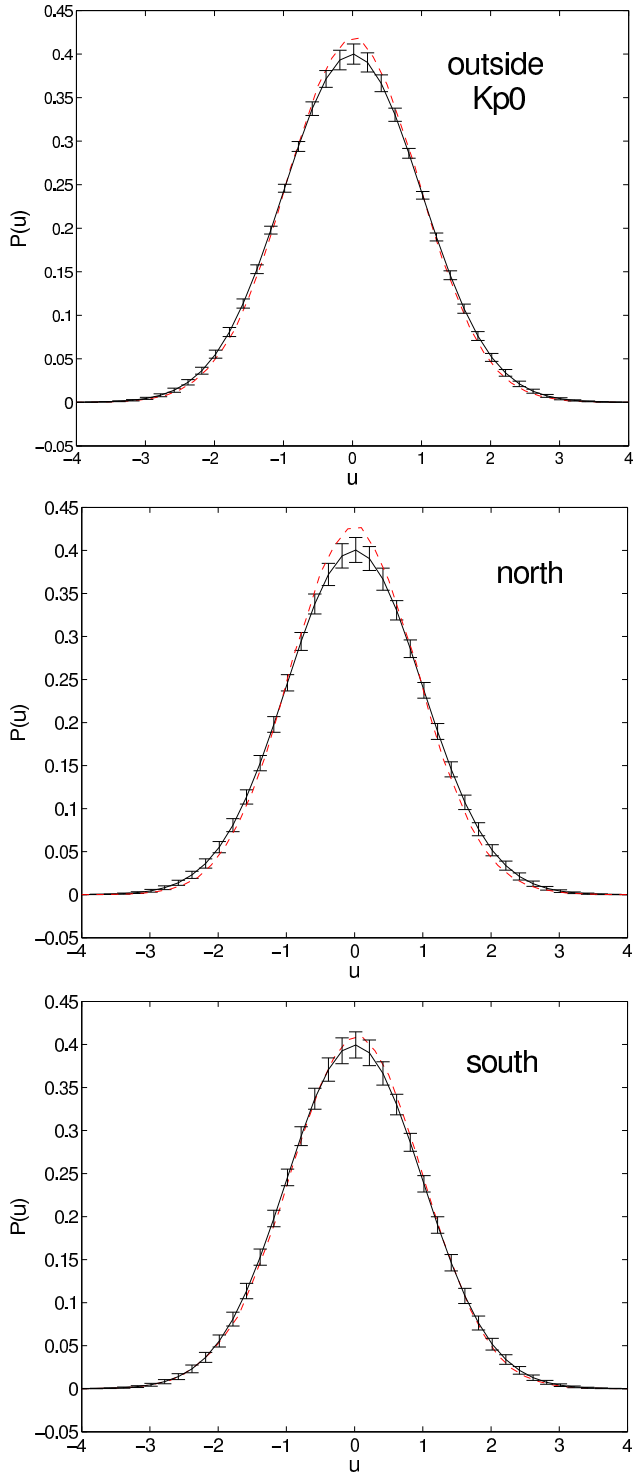}
\caption{The top panel shows the histogram of the normalised WMAP data
outside the Kp0 mask (dashed red line) compared to the averaged
histogram obtained from 1000 simulations (solid black line). The
error bars indicate the dispersion obtained from simulations. The
middle and bottom panels give the same histograms for the northern and
southern ecliptic hemispheres (only considering pixels outside the
Kp0) respectively.}
\label{fig:hist_norm}
\end{center} 
\end{figure}

We have calculated different estimators for the normalised data
outside the Kp0 mask: dispersion, skewness and kurtosis. In addition,
the data have been compared to a N(0,1) distribution through a
Kolmogorov-Smirnov test. In order to assign a significance to these
quantities, the same analysis has been applied to 1000 Gaussian
simulations of the WMAP combined data. To reproduce the data as
closely as possible, the simulations have been generated at resolution
$n_{side}$=512 for each radiometer and then processed in the same way as
the data to obtain the simulated combined maps. Results are shown
in Table \ref{tab:1pdf_3yr_results}.
\begin{table}
 \begin{center} 
  \caption{\label{tab:1pdf_3yr_results} The dispersion,
  skewness, kurtosis and Kolmogorov-Smirnov distance d$_{\rm KS}$ are
  given for the normalised WMAP data as well as for the northern and
  southern ecliptic hemispheres (in all cases considering only pixels
  outside the Kp0 mask). The significance of each quantity has been
  calculated from 1000 Gaussian realisations as the percentage of
  simulations with a value larger or equal to that obtained from
  the data.}
  \begin{tabular}{ccc}
  \hline
Estimator & WMAP value & Signif. ($\%$) \\
  \hline
& All pixels outside Kp0 mask& \\
  \hline
 dispersion & $9.55 \times 10^{-1}$ & $97.8$   \\ 
 skewness & $-1.91 \times 10^{-2}$~~~ & $74.7$   \\ 
 kurtosis & $1.49 \times 10^{-2}$ & $20.9$   \\ 
 d$_{\rm KS}$ & $1.21 \times 10^{-2}$ & $7.0$   \\ 
  \hline
 & Northern ecliptic hemisphere &  \\
  \hline 
  dispersion & $9.35 \times 10^{-1}$ & $99.4$   \\ 
  skewness & $2.17 \times 10^{-2}$ & $28.1$   \\ 
  kurtosis & $2.18 \times 10^{-2}$ & $20.1$   \\ 
  d$_{\rm KS}$ & $1.89 \times 10^{-2}$ & $25.2$   \\ 
  \hline
 & Southern ecliptic hemisphere &  \\
  \hline
  dispersion & $9.76 \times 10^{-1}$ & $71.8$   \\ 
  skewness & $-5.85 \times 10^{-2}$~~~ & $92.8$   \\ 
  kurtosis & $7.08 \times 10^{-4}$ & $35.4$   \\ 
  d$_{\rm KS}$ & $1.05 \times 10^{-2}$ & $67.3$   \\ 
  \hline
  \end{tabular}
 \end{center}
\end{table}
We find that the skewness and kurtosis are consistent with the results
obtained from Gaussian simulations. However, the dispersion of the
normalised data is significantly lower than expected, at the level of
97.8 per cent (i.e., only 22 out of the 1000 simulations have lower
values than the one found for the data). Regarding the
Kolmogorov-Smirnov distance (d$_{\rm KS}$), we find that only 7 per
cent of the simulations have a larger value of d$_{\rm KS}$ than the
data. This could be related to the fact that the data dispersion is
lower than expected, what would tend to produce a deviation of the
normalised temperature from a N(0,1) distribution. These deviations
can also be seen in the top panel of Figure~\ref{fig:hist_norm}, which
compares the histogram obtained from the normalised WMAP data with the
one obtained averaging over 1000 Gaussian simulations. A deviation of
the data histogram (dashed line) from the one obtained with
simulations (solid line) is apparent due to the small value of the
dispersion of the normalised data. We would like to mention that the
same analysis has also been performed on the first year WMAP data,
finding again a low value of the dispersion (at a significance level
of 99.5 per cent).

We may wonder if errors in the estimation of the WMAP noise level
could affect the results of table~\ref{tab:1pdf_3yr_results}. In
particular, taking into account equation (\ref{eq:norm_temp}), an
overestimation of the noise level would tend to bias the dispersion of
the normalised temperature towards values lower than
unity. \cite{jar07} use two different methods to estimate the noise
level of each WMAP radiometer, finding that both methods agree within
0.3 per cent. This value can be considered as an indication of the
level of error expected in the estimation of the WMAP noise. Even if
we consider an overestimation of 1 per cent in the noise dispersion
and recalculate, accordingly, the normalised temperature for the data,
we still find a significant deviation for the dispersion. In
particular, the new value of the dispersion is 0.956, which
corresponds to a significance of 97.4 per cent. Therefore, the level
of errors expected in the estimation of the noise level cannot
explain this anomaly.

In order to investigate further the origin of this deviation, we have
repeated the same analysis in the northern and southern ecliptic
hemispheres, considering only those pixels outside the Kp0 mask (which
correspond to a sky fraction of 38.8 and 37.0 per cent for the
northern and southern hemispheres, respectively).  Results are shown
in Table~\ref{tab:1pdf_3yr_results}.  The values of skewness, kurtosis
and d$_{\rm KS}$ remain compatible with Gaussian simulations. However,
we find an asymmetry in the behaviour of the dispersion: while the
data dispersion in the southern ecliptic hemisphere is compatible with
the simulations (at the 71.8 per cent level), a clear deviation
is found in the north, at the 99.4 per cent significance level (i.e.,
only 6 out of 1000 simulations had lower dispersion values). The
middle (bottom) panel of Figure~\ref{fig:hist_norm} shows the
histogram of the data in the northern (southern) ecliptic hemisphere
compared to the averaged histogram obtained from Gaussian simulations
in the same sky region. The histogram obtained from the northern data
shows again a systematic deviation from the average one, which is even
larger than in the case of the whole sky, due to the lower value of
the dispersion found for the data of the northern ecliptic
hemisphere. Conversely, the histogram obtained from the southern data
is closer to the average value of the Gaussian simulations.

\section{Study of the CMB dispersion $\sigma_{0}$ } 
\label{sec:sigma0}

Given the deviation found in the dispersion of the normalised
temperature, we have performed a more detailed study of the CMB
dispersion $\sigma_0$. Let us recall that the dispersion of the data
$\sigma(\mathbf{x})$ has contributions from both the CMB signal and the
instrumental noise:
\begin{equation}
\sigma(\mathbf{x})=\sqrt{\sigma_0^2 + \sigma_n^2(\mathbf{x})}
\label{eq:data_disp}
\end{equation}
We have estimated the value of $\sigma_0$ by allowing this quantity to
vary in equation (\ref{eq:norm_temp}) and compare the different
obtained normalised temperatures to a Gaussian of zero mean and unit
dispersion with the Kolmogorov-Smirnov test. Our estimation of the CMB
dispersion is given by the value of $\sigma_0$ that produces the
minimum d$_{\rm KS}$. We have tested with simulations that this
estimator performs well, as it will be shown below for the different
considered cases.

\begin{figure}
\begin{center}
\includegraphics[angle=0,width=\hsize]{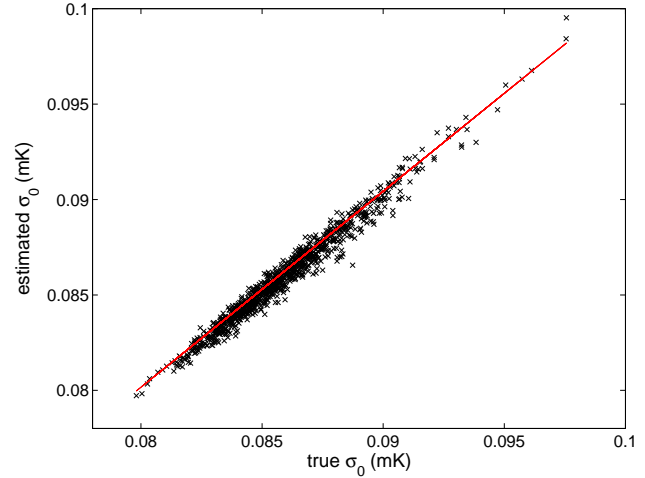} 
\caption{Estimated versus true dispersion (in mK) obtained from 1000
simulations of the WMAP combined map outside the Kp0 mask. For
illustration, the best fit to a straight line is also shown (solid red
line), which corresponds to $\hat{\sigma}_0=1.027 \sigma_0 - 0.002$.}
\label{fig:estim_linear}
\end{center} 
\end{figure}

Using this method, we have estimated a value of $\sigma_0 = 8.16
\times 10^{-2}$ mK for the WMAP combined map. The bias and error of
this estimation has been tested using 1000 CMB simulations.
Figure~\ref{fig:estim_linear} shows the true CMB dispersion of each
simulation outside the Kp0 mask versus the one estimated with our
method from the simulated data in the same region of the sky. For
illustration, the best fit to a straight line is also plotted, showing
a very good correlation between the true and estimated
dispersions. The bias (obtained as the average difference of the true
minus the estimated dispersion) and error (obtained as the dispersion
of the same difference) of the estimator are $b=-6.9 \times 10^{-5}$
mK and $e=4.2 \times 10^{-4}$ mK respectively, i.e. the error of our
estimator is at the level of $\sim$0.5 per cent. In addition, the bias
is also significantly smaller than the typical value expected for the
CMB dispersion and therefore our estimator can be considered unbiased
to a very good approximation. An additional bias in the estimated
dispersion could also be present if the assumed level of instrumental
noise were incorrect. In particular, an overestimation of the noise
would imply an underestimation of the CMB dispersion. To check this
possibility, we have also estimated $\sigma_0$ as $\left<T_1 T_2
\right> $, with $T_1 = Q1+V1+W1+W2$ and $T_2 = Q2+V2+W3+W4$, finding a
value which is consistent with the previous estimation within the
1$\sigma$ error. Note that since instrumental noise is uncorrelated
between channels, this latter estimation is not affected by a possible
bias in the noise level.

For comparison, we have also calculated the mean $\sigma_0$ from 20000
noiseless CMB Gaussian simulations of the WMAP combined data, finding
a value of $8.60 \times 10^{-2}$ mK, clearly higher than the
dispersion estimated for the data. In order to quantify the
significance of this result, we have also obtained the distribution of
variances from the same simulations. We find that 98.7 per cent of the
simulations have a larger value of the variance (or, equivalently, of
the dispersion) than the one found for the WMAP data. These results
are summarised in Table~\ref{tab:nseh_combined_3yr}.
\begin{table*}
 \begin{center} \caption{\label{tab:nseh_combined_3yr} The second
  column gives the CMB dispersion (in mK) estimated for the WMAP
  combined data obtained using pixels of three different regions of
  the sky: outside the Kp0, the northern ecliptic hemisphere outside
  the Kp0 and the southern ecliptic hemisphere outside the Kp0. In the
  third column, the mean dispersion value of the model, obtained
  averaging over 20000 Gaussian simulations, is given. Significances,
  that have been calculated as the percentage of simulations with
  values larger than the one found for the data using 20000 Gaussian
  CMB simulations, are given in the fourth column. Finally, in the
  last two columns, the bias (obtained as the average differences of
  the true minus the estimated dispersion) and error of the estimator,
  both in mK, are given, which have been obtained from 1000 Gaussian
  simulations.}
  \begin{tabular}{c c c c c c}
  \hline
Region & $\hat{\sigma}_{0}$ & $\sigma_0^{model}$ & Signif. & Bias & Error  \\
  \hline
  Kp0 & $8.16 \times 10^{-2}$ & $8.60 \times 10^{-2}$ & $98.7$ & $-6.9
  \times 10^{-5}$ & $4.2 \times 10^{-4}$  \\  
  north & $7.97 \times 10^{-2}$ &$8.59 \times 10^{-2}$ & $99.8$ &
  $-1.6 \times 10^{-4}$ & $6.8 \times 10^{-4}$  \\  
  south & $8.37 \times 10^{-2}$ &$8.59 \times 10^{-2}$ & $75.7$ &
  $-2.0 \times 10^{-4}$ & $7.2 \times 10^{-4}$  \\
  \hline 
  \end{tabular}
 \end{center}
\end{table*}

It is interesting to point out that, for a Gaussian CMB, the
distribution of the correlation function $C(\theta)$ for a given power
spectrum can be analytically calculated \citep{cay91}. In particular,
we are interested in obtaining the theoretical distribution of the
variance, i.e., the correlation function at $\theta=0$. Following
\cite{cay91}, the cumulative function of the variance can be
calculated as
\begin{eqnarray}
F(\sigma_0^2) &=& \frac{1}{2} + \frac{1}{\pi}\int_{0}^{\infty}
\prod_{\ell =0}^{\ell_{max}}{\left[ \left( 1 + 4 t^{2}
\sigma_{\ell}^{2} \right)^{-\frac{2 \ell +1}{4}} \right]} \nonumber \\
&&\frac{\sin{\left[B\left(t \right) + t \sigma_0^2 \right]}}{t} 
\it{dt} \nonumber \\
B \left( t \right) &=& \sum_{\ell=0}^{\ell_{max}}
\frac{2 \ell +1}{2} \arctan{\left(-2 t \sigma_{\ell} \right)} ~,
\label{eq:signif_laura}
\end{eqnarray}
where $\sigma_{\ell}=\frac{C_{\ell}}{4\pi}$ contains the power
spectrum dependence. Calculating the derivative of the cumulative
function with respect to $\sigma_0^2$, we can also obtain the probability
distribution of the variance:
\begin{equation}
p(\sigma_0^2)= \frac{1}{\pi}\int_{0}^{\infty}
\prod_{\ell =0}^{\ell_{max}}{\left[ \left( 1 + 4 t^{2}
\sigma_{\ell}^{2} \right)^{-\frac{2 \ell +1}{4}} \right]} \cos{\left[B\left(t \right) + t \sigma_0^2 \right]} \it{dt}
\label{eq:pdf_laura}
\end{equation}
\begin{figure*}
\begin{center}
\includegraphics[angle=0, width=\hsize]{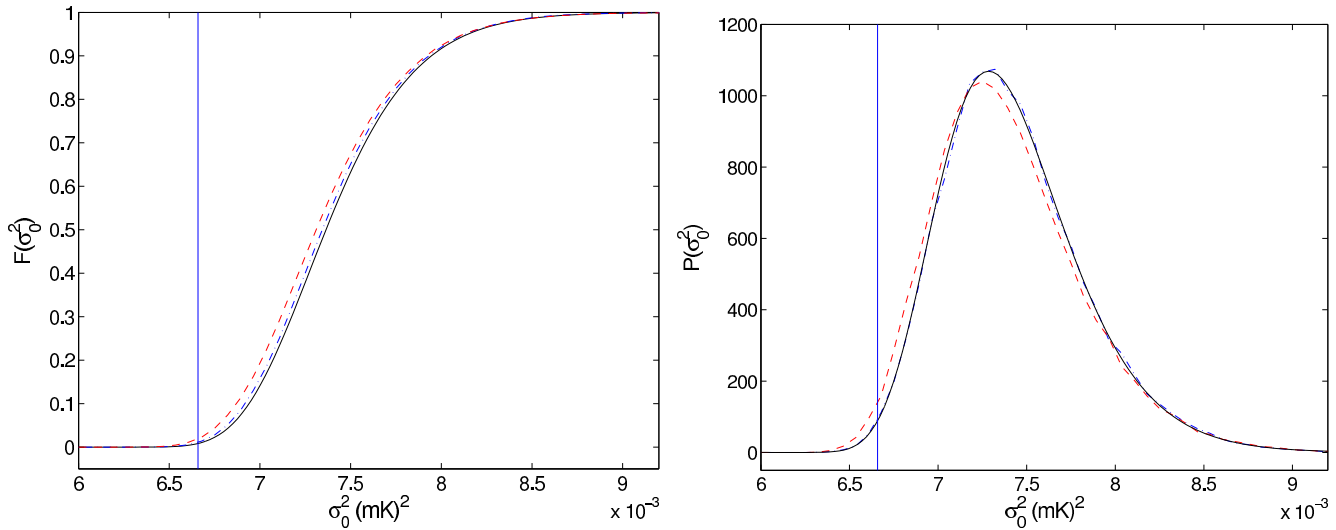}
\caption{Left panel: theoretical cumulative function of the CMB
variance (solid black line), compared to the same
functions obtained from 60000 Gaussian simulations over the whole sky
(blue dotted-dashed line) and using only the pixels outside the Kp0
mask (red dashed line). Right panel: the probability distribution of
the variance is given for the three same cases. For illustration, the
solid vertical line in both panels indicates the value obtained from
the WMAP data. The units of the variance are $(mK)^2$.}
\label{fig:pdf_variances}
\end{center} 
\end{figure*}
The previous distributions are valid when considering the variance of
the whole sky. However, the presence of the Kp0 mask restricts the
fraction of the sky ($\sim$ 76 per cent) used in the analysis. Since
obtaining the significance of a given estimated CMB variance from
simulations is significantly slower than calculating it from the
cumulative function (\ref{eq:signif_laura}), we have tested whether
the theoretical distributions are a good approximation in our
case. The left panel of Figure~\ref{fig:pdf_variances} shows the
cumulative function of $\sigma_0^2$ (black solid line) obtained using
the theoretical power spectrum of the WMAP combined map calculated in
Appendix~\ref{ap:cl_comb_teo}. The dotted-dashed blue line corresponds
to the distribution of the variance obtained from 60000 Gaussian
simulations over the whole sky. As expected, the agreement in this
case is very good. Finally, the red dashed line gives the distribution
of the variance obtained from the same simulations considering only
those pixels outside the Kp0 mask. We can appreciate some small
differences between this curve and the one obtained from equation
(\ref{eq:signif_laura}), due to the fact that we are applying a mask,
but the general agreement is very good. The maximum difference between
both cumulative functions is found in the central part of the plot and
is less than 0.04, whereas the differences in the tails of the
distributions are even smaller. Therefore, the theoretical cumulative
function provides a simple and good approximation to obtain the
significance of the estimated CMB variance in the considered case.  In
particular, using the theoretical distribution we obtain a
significance of 99.3 per cent for the estimated $\sigma_0$, that
should be compared with the value of 98.7 obtained from simulations.
Similar conclusions can be derived from the right panel of
Figure~\ref{fig:pdf_variances}, that shows the probability distribution
of the variance for the same cases as before: theoretical (solid black
line), from simulations using the whole sky (blue dotted-dashed line)
and from simulations considering pixels outside Kp0 (red dashed-line).

Taking into account the results found in the previous section, we have
also estimated the CMB dispersion in the northern and southern
ecliptic hemispheres (considering only those pixels outside the Kp0
mask), obtaining values of 7.97$\times 10^{-2}$ and 8.37$\times
10^{-2}$ mK, respectively. The corresponding biases, errors, mean
dispersion of the model and significances are given in
Table~\ref{tab:nseh_combined_3yr}. As expected, the bias and error of
the estimator has increased slightly with respect to the ones obtained
using all pixels outside the Kp0 mask, due to the smaller fraction of
sky considered, but they are still small (less than 1 per cent with
respect to $\hat{\sigma}_0$). In addition, it should be noted that
this bias is negative, implying that, if anything, the CMB dispersion
would tend to be overestimated and, therefore, the true $\sigma_0$
would be even lower.

We remark that the variance of the northern hemisphere is
significantly low (only 0.2 per cent of the Gaussian simulations have
lower values) whereas the southern hemisphere is compatible with the
simulations. This is consistent with the results found in
\S~\ref{sec:tests}, which seem to indicate that the anomaly found in
the WMAP variance comes from the northern ecliptic hemisphere. This
result is also in agreement with the lack of power found in the
northern ecliptic hemisphere of the WMAP data by previous works
\citep{eri04,han04a,eri05}.

We have also studied the ratio of $\sigma_0^2$ in the northern
ecliptic hemisphere over $\sigma_0^2$ in the southern one for the WMAP
data, finding that this quantity is consistent with what is expected
from Gaussian simulations (with a significance of 88.7 per cent
obtained from 20000 simulations). This can be explained because,
although the value of $\sigma_0$ in the southern hemisphere is
consistent with Gaussianity, this quantity is lower than the average
dispersion expected from simulations, and this fact makes the
north-south ratio compatible with the Gaussian model.

Finally, we have studied the effect that the low CMB
quadrupole measured by COBE \citep{benn96} and WMAP
\citep{benn03a,hin07} has in our results. Taking into account equation
(\ref{eq:cmb_disp}) and the fact that the $C_{\ell}$'s are roughly
proportional to $1/\ell (\ell + 1)$, it is easy to see that
the largest contributions to the CMB variance come from the lower
multipoles and, in particular, from the quadrupole. In order to test
if the low CMB quadrupole can be the origin of the low CMB variance we
have repeated our analysis after removing the best-fit quadrupole
(outside the Kp0 mask) from the data. With this procedure we find
$\hat{\sigma}_0 = 8.07 \times 10^{-2}$ mK which corresponds to a
theoretical significance of 96.9 per cent (obtained using equation
(\ref{eq:signif_laura}) setting $C_2=0$). In addition, we have obtained the
significance from 20000 simulations, where the quadrupole has been
removed, finding a significance of 95.0 per cent. The same analysis
has also been performed in the northern and southern ecliptic
hemispheres, estimating for the dispersion $7.91 \times
10^{-2}$ mK and $8.30 \times 10^{-2}$ mK, respectively. The
significances of these values, obtained from 20000 Gaussian
simulations, are 99.2 and 44.3 per cent, respectively. Therefore,
after removing the quadrupole, although the significances are slightly
reduced, the CMB variance of the whole sky and that of the northern ecliptic
hemisphere are still anomalously low. Thus, the low quadrupole cannot
explain by itself these deviations.

\section{Discussion}  
\label{sec:4}

We may wonder if the anomalous value found for the CMB dispersion in
the WMAP data may have an extrinsic origin, i.e., to be due to the
presence of some systematics or foreground residuals. In order to
clarify this point, we have performed some further tests including an
analysis of each individual radiometer (\S
\ref{sec:single}) as well as the study of the effect of 1/f noise (\S
\ref{sec:1/f_noise}) and foreground residuals (\S \ref{sec:fore}).

\subsection{Single radiometer analysis}
\label{sec:single}

In order to check if the low value of $\sigma_0$ found in the WMAP
data could be explained by the presence of an anomalous radiometer, we
have estimated the CMB dispersion from the maps of each single
radiometer following the method explained in \S\ref{sec:sigma0}. The
results are summarised in Table~\ref{tab:bs0_radiom}. The mean
dispersion $\sigma_0^{model}$ and the significances (see \S\ref{sec:sigma0})
have been obtained analytically using the power spectrum of the WMAP
best-fit model \citep{spe07} and taking into account the different
beam window functions of each radiometer as well as the window pixel
function of the considered resolution ($N_{side}=256$). The biases and
errors of the estimator have been obtained from 100 Gaussian
simulations and are at a level less than 1 per cent in all cases.
\begin{table*}
 \begin{center} \caption{\label{tab:bs0_radiom} CMB dispersion (in mK)
  estimated from maps of individual radiometers outside the Kp0
  mask. The mean dispersion of the model and the significances have
  been obtained analytically. The corresponding bias and error (in mK)
  have been obtained from 100 Gaussian simulations. For comparison,
  the same results obtained for the WMAP combined map are also
  shown. An estimation of the $\sigma_0$ of the combined map from each
  of the radiometers and its corresponding error are also given in the
  last two columns (see text for details). }
  \begin{tabular}{c c c c c c c c}
  \hline
Radiometer & $\hat{\sigma}_{0}$ & $\sigma_0^{model}$  & Signif. ($\%$)
& Bias & Error &  
  $\hat{\sigma}_{0} + d$ & Error (d)\\
  \hline
Q1 & $7.93 \times 10^{-2}$ & $8.34 \times 10^{-2}$  & $98.0$ & $-4.9
\times 10^{-5}$ & $4.1 \times 10^{-4}$ & $8.21 \times 10^{-2}$ & $3.7\times 10^{-4}$\\ 
Q2 & $7.88 \times 10^{-2}$ & $8.29 \times 10^{-2}$  & $97.9$ & $ 4.1 \times 10^{-6}$ & $4.5 \times 10^{-4}$ & $8.21 \times 10^{-2}$ & $3.7\times 10^{-4}$\\  
V1 & $8.42 \times 10^{-2}$ & $8.84 \times 10^{-2}$  & $98.9$ & $ 1.0 \times 10^{-5}$ & $4.4 \times 10^{-4}$ &$8.19 \times 10^{-2}$ & $3.7\times 10^{-4}$\\  
V2 & $8.46 \times 10^{-2}$ & $8.88 \times 10^{-2}$  & $98.8$ & $-8.4 \times 10^{-6}$ & $3.7 \times 10^{-4}$ &$8.19 \times 10^{-2}$ & $3.6\times 10^{-4}$\\   
W1 & $8.66 \times 10^{-2}$ & $9.05 \times 10^{-2}$ & $98.3$ & $-2.5 \times 10^{-5}$ & $5.4 \times 10^{-4}$ &$8.20 \times 10^{-2}$ & $5.7\times 10^{-4}$\\  
W2 & $8.48 \times 10^{-2}$ & $8.87 \times 10^{-2}$ & $98.1$ & $-8.7 \times 10^{-6}$ & $5.8 \times 10^{-4}$ &$8.21 \times 10^{-2}$ & $6.6\times 10^{-4}$\\  
W3 & $8.47 \times 10^{-2}$ & $8.90 \times 10^{-2}$ & $99.1$ & $ 1.1 \times 10^{-4}$ & $6.8 \times 10^{-4}$ &$8.16 \times 10^{-2}$ & $6.5\times 10^{-4}$\\  
W4 & $8.74 \times 10^{-2}$ & $9.07 \times 10^{-2}$ & $95.9$ & $ 8.2 \times 10^{-5}$ & $5.8  \times 10^{-4}$ & $8.26 \times 10^{-2}$ & $5.9\times 10^{-4}$\\
\hline
Combined & $8.16 \times 10^{-2}$ & $8.62 \times 10^{-2}$ & $99.3$ & $-6.9 \times 10^{-5}$ & $4.2 \times 10^{-4}$ && \\ 
  \hline
  \end{tabular}
 \end{center}
\end{table*}

As seen in Table \ref{tab:bs0_radiom}, a different value of $\sigma_0$
has been obtained for each radiometer, which is expected due to the
different beam window functions of each radiometer as well as to the
statistical error of our estimator. In order to check if the
dispersion of each radiometer is consistent with each other, and that
no single radiometer has an anomalous dispersion that could be
responsible for the low value of $\sigma_0$ found in the combined map,
we have performed the following test. First, we have obtained the
mean, $d$, and dispersion, error($d$), of the difference between the
dispersion of the combined map minus that of each single radiometer
from 100 simulations, where the map dispersions were obtained using
our estimator. We have then added the corresponding mean difference
$d$ to the $\hat{\sigma}_0$ of each radiometer, which provides an
estimation of $\sigma_0$ for the combined map (given in the sixth
column of Table~\ref{tab:bs0_radiom}). If all the radiometer maps are
consistent with each other, all these quantities should be compatible
among them within the range allowed by the dispersion of the
calculated differences (given in the seventh column of
Table~\ref{tab:bs0_radiom}) as well as with the $\hat{\sigma}_0$
derived for the WMAP combined data. From Table~\ref{tab:bs0_radiom},
we see that consistency is found and that no single radiometer has an
anomalous dispersion.\footnote{Although consistency is found, it
is observed that the dispersions derived from the radiometers are in
general higher than that derived directly from the WMAP combined
map. This small bias could be due to the presence of residual
foregrounds and 1/f noise in the single radiometer maps, whose effect,
as will be shown in sections \S\ref{sec:1/f_noise} and
\S\ref{sec:fore}, is to produce an overestimation of the dispersion in
these maps. This is consistent with the fact that the radiometers that
have a dispersion that deviates more from the one of the combined map
(although always within the 2$\sigma$ error) are Q1 and Q2, which are
expected to be the channels more affected by residuals foregrounds,
and W4, which presents the largest 1/f noise of all the radiometers
\citep{jar03}. These possible residuals are expected to be reduced,
relatively to the CMB signal, in the combined map and, therefore, the
direct estimation of the $\sigma_0$ from the combined data would tend
to give a lower (less biased) value than that obtained from each of
the radiometers.} Moreover, the significances obtained for the
estimated dispersion for all the radiometers are significantly high,
confirming that the anomaly cannot be attributed to a single
radiometer.

In addition, we have repeated the same analysis for all the
radiometers for each individual year of data. We find that the estimated CMB
dispersion is reasonably stable along the three years of data for all the
radiometers as well as for the combined map. This shows that the
low $\sigma_0$ cannot be explained as the result of some anomaly in
a single year of data.

\subsection{1/f noise}
\label{sec:1/f_noise}

In our analysis, it has been assumed that the WMAP instrumental
noise is a Gaussian white noise characterised by a dispersion
$\sigma_n(\mathbf{x})$ which varies with the position of the
sky. Although this seems to be a very good approximation, the WMAP
data also contain a small level of 1/f noise \citep{jar03,jar07} 
and therefore we may wonder if this could be affecting our estimation
of $\sigma_0$.

In order to test the effect of 1/f noise we need realistic noise
simulations that reproduce the WMAP data pipeline. As far as we know,
this type of simulations are not available for the three-year data
but, however, the WMAP team did provide 110 of these simulations for
each radiometer for the first year of data that include 1/f noise as
well as all known radiometric effects (for details see the information
in the LAMBDA web site).


Therefore, we have carried out the following test. First of all, we
have combined the simulations of the different radiometers in the
appropriate way in order to obtain 110 realistic noise simulations of
the first-year WMAP combined map. We have then obtained the realistic
noise dispersion map from these simulations. Since the number of
simulations is relatively small, if we try to obtain the dispersion at
each pixel, this map will be very noisy. Therefore, we have assumed that
this quantity varies smoothly over the map and that neighbouring
pixels would have the same level of noise. In particular, we have
obtained the dispersion from 16 neighbouring pixels along the 110
simulations. The mean value of the ratio of this dispersion map over
the one constructed assuming that only white noise is present is 1.02,
which is an indication of the small level of 1/f noise present in the
data.

The next step is to construct a new set of 110 realistic simulations
by adding the combined realistic noise simulations to Gaussian CMB
realizations of the WMAP combined first-year data. We have then
applied our dispersion estimator to this new set of simulations and
study its bias and error in 2 different cases: 1) including in
equation (\ref{eq:data_disp}) the value of $\sigma_n(\mathbf{x})$
obtained from the realistic noise simulations and 2) assuming that we
have only white noise. Interestingly, the errors in both cases are
very similar ($\sim 4 \times 10^{-4}$ mK),
but the biases are significantly different. In the first case, when
the noise is correctly characterised, the bias is very small ($-3.4
\times 10^{-5}$ mK), however, when only white noise is erroneously
assumed, the bias increases to a value of $-4.6 \times 10^{-4}$
mK. This means that, if 1/f noise were present in the data and we did
not properly take it into account, our estimation of the CMB
dispersion would tend to shift towards higher values and, therefore,
this effect could not explain an anomalous low value of
$\sigma_0$. This can be understood, since, by not including the 1/f
noise, we are underestimating the level of instrumental noise and
this extra power would tend to be compensated by increasing the signal
dispersion. As a further test, we have estimated the CMB dispersion
from the WMAP first-year data using the realistic noise dispersion
map, finding a value of $8.12 \times 10^{-2}$ mK to be compared with
the larger dispersion obtained assuming white noise ($8.17 \times
10^{-2}$ mK).

Since there are not realistic noise simulations available for the WMAP
three-year data, we cannot repeat this test for the complete set of
data. However, we can qualitatively check that the same argument is
valid for the three-year data using the reduced resolution inverse
noise covariance matrices provided in the LAMBDA web site for the one
and three-year data. The diagonal elements of the (direct) noise
covariance matrices are dominated by white noise, whereas the
off-diagonal elements come from 1/f noise. We have obtained the ratio
of the diagonal and off-diagonal terms of the first-year data over the
corresponding elements of the three-year data finding that, in both
cases, this ratio peaks around 3. This means that the relative importance of
the 1/f noise with respect to the white noise is approximately the
same in both sets of data and, therefore, the effect that the 1/f
noise has in the three-year data should be very similar to the one
found for the first-year data. 

Thus, residual 1/f noise cannot explain the low CMB variance found
for the three-year WMAP data, since its unaccounted presence, if
anything, would lead to higher values of $\hat{\sigma}_0$.

\subsection{Foreground residuals}
\label{sec:fore}

Although we are using the foreground cleaned maps provided by the WMAP
team and applying a Galactic plus point source mask, some foreground
residuals may still be present in the data. Therefore, it is
interesting to test if this possible contamination can affect our
estimation of the CMB dispersion.

An indication of a significant effect from foregrounds would be that
the anomaly found in $\sigma_0$ was frequency dependent. However, from
Table~\ref{tab:bs0_radiom}, we can see that this is not the case,
since $\sigma_0$ is consistently low in the whole range of frequencies
probed by Q, V and W (from 41 to 94 GHz).

In order to study further the effect of foreground contamination, we
have estimated the value of $\sigma_0$ in the combined and individual
radiometer maps without the foreground correction. The most relevant
change is found for the Q1 and Q2 maps (lowering the significances of
$\sigma_0$ down to 90.7 and 86.2 per cent respectively), whereas for
the radiometers of the V and W bands, the changes are smaller. This is
consistent with the fact that foreground emission is higher in the Q
band and therefore the cleaning of the Q1 and Q2 maps is important to
improve the estimation of the CMB dispersion. The presence of
residuals also translates into a larger value for the $\hat{\sigma}_0$
of the combined map ($8.23 \times 10^{-2}$ mK).

As one would expect, these results indicate that if foreground
residuals were present in the data, this would tend to bias
$\sigma_0$ towards higher values. Therefore, the presence of
foreground contamination cannot explain the anomaly found for the CMB
dispersion.


\section{Possible explanations of the anomaly}
\label{sec:explanations}

In all the analyses carried out in this work, we have assumed
that the CMB fluctuations are isotropic and follow a Gaussian
distribution characterised by a given power spectrum (which in turn is
described by a particular cosmological model). However, since we have
found an anomaly regarding the variance that cannot be easily
explained by foregrounds or systematics, we may wonder if some of
these assumptions are incorrect. In \S\ref{sec:parameters} we discuss
if a different choice of cosmological parameters can make the
estimated variance consistent with the model. In \S\ref{sec:isotropy}
we study if this anomaly can be due to a more fundamental reason: a
deviation of the CMB from Gaussianity and/or isotropy.

\subsection{Effect of cosmological parameters}
\label{sec:parameters}
During this work, we have used for the CMB power spectrum the
best-fit model to the WMAP data assuming a flat $\Lambda$CDM model
\citep{spe07}. However, we would like to test if by varying the values of
the cosmological parameters -- within the current uncertainties --
the estimated CMB variance could be more consistent with the
theoretical expectations.

In order to test this possibility, we have considered the effect of
varying the following cosmological parameters: the physical baryon
density $\Omega_{b}h^{2}$, the physical cold dark matter density
$\Omega_{c}h^{2}$, the reduced Hubble constant $h$, the reionization
optical depth $\tau$, the spectral index $n_{s}$ and the amplitude of
the density fluctuations $A$ (with the constraint of having a flat
$\Lambda$CDM model). Given that we have only one observable, the CMB
variance $\sigma_0^2$, modifying all the cosmological parameters
simultaneously would lead to strong degeneracies between
them. Therefore, as a first approach, we have studied the effect of
varying one single parameter (within a range given by 2 times the
errors of the WMAP best-fit model) while leaving the others
unchanged. In particular, we have constructed the corresponding
one-dimensional likelihoods in the following way: for each set of
parameters, the CMB spectrum was calculated using CAMB \citep*{lew00};
this power spectrum was then inserted in equation (\ref{eq:cl_comb3})
in order to get the $C_{\ell}$'s of the WMAP combined map; finally,
the probability of having the observed value of $\sigma_0^2$ for that
power spectrum was obtained using equation (\ref{eq:pdf_laura}).

\begin{table}
 \begin{center} \caption{\label{tab:parameters} Ratio between
  the maximum value of the likelihood, when varying each parameter
  within a range given by 2 times the errors of the WMAP best fit
  model, over the likelihood of the best fit model (column 2) and 
  significance of the variance for the model correspoding to this
  maximum likelihood value (column 3). }
  \begin{tabular}{c c c }
  \hline
Parameters & Ratio & Signif.(\%) \\
  \hline
$\Omega_{b}h^2$ & 2.1 & 98.0 \\
$\Omega_{c}h^2$ & 6.8 & 90.3 \\
$h$ & 3.6 & 96.5 \\
$\tau $ & 10.1 & 68.5 \\
$n_s$ & 11.9 & 79.0 \\
$A$ & 14.1 & 57.5 \\
  \hline
  \end{tabular}
 \end{center}
\end{table}

We find that the estimated CMB variance favours lower values of
$\Omega_{b}h^{2}$, $h$ and $A$ than the ones of the best-fit model
whereas higher values of $\Omega_{c}h^{2}$, $n_{s}$ and $\tau$ are
preferred. However, not all parameters are equally sensitive to the
variance as can be seen in Table~\ref{tab:parameters}. This table
gives the ratio between the maximum value of the likelihood, when
varying one parameter in the considered range, over the likelihood of
the WMAP best-fit model. The higher this ratio, the more sensitive the
parameter to the CMB variance. We also show in
Table~\ref{tab:parameters} the significance of the variance for the
model correspoding to this maximum likelihood value.  We see that the
compatibility of $\sigma_0$ with the model is only moderately improved
when varying $\Omega_{b}h^{2}$, $h$ and $\Omega_{c}h^{2}$ within the
allowed range and the likelihood ratio also confirms that these
parameters are not very sensitive to the value of the CMB
variance. However, this ratio reaches higher values when considering
the variation of $A$, $\tau$ and $n_s$ and also the $\sigma_0$ becomes
perfectly compatible with the model, showing that these three
parameters are more relevant regarding this observable.

Taking into account the previous results, we have further investigated
the effect of cosmological parameters by constructing the likelihood
varying simultaneously the three more sensitive parameters,
i.e. $\tau$, $n_{s}$ and $A$, while fixing the others to the values
of the best-fit model. We find that there is a strong degeneracy
between $\tau$ and $A$, and, therefore, we cannot extract any further
information with the observed CMB variance regarding these two
parameters. However, the situation is different for the spectral
index. We have constructed the marginalized likelihood of $n_s$,
finding that values of the spectral index higher than the ones given
by the WMAP best-fit model are favoured. In particular, we find
$\langle n_{s} \rangle= 0.98$, although one should point out that, as
expected since we are using only one observable, the likelihood
constructed in this way is significantly wider than the one obtained
with the full power spectrum of the WMAP data.

Therefore it is possible to find models consistent with the
estimated CMB variance by varying the value of the cosmological
parameters in a certain range. However, in order for a model to be
valid, it should still be compatible with the measured power
spectrum. Thus, in principle, we should look for the model that best fits the
data, allowing to vary all the cosmological parameters at the same
time, subject to the constraint of compatibility with the estimated
CMB variance. Unfortunately, in practice, this task is very
complicated, and, as an illustration, we have just tried to find one model
(not necessarily the best possible) that fits the WMAP power spectrum
and is also consistent with the estimated CMB variance.

\begin{table}
 \begin{center} \caption{\label{tab:models} 
Estimated cosmological parameters for the WMAP best-fit and
for an illustrative model more consistent with the estimated variance.}
  \begin{tabular}{c c c }
  \hline
Parameters & Best-fit & New model \\
  \hline
$100\Omega_{b}h^2$ & $2.23^{+0.07}_{-0.09}$ & 2.27 \\
$\Omega_{c}h^2$ & $0.104^{+0.007}_{-0.010}$& 0.089 \\
$h$ & $0.73^{+0.03}_{-0.04}$ & 0.81 \\
$\tau $ & $0.088^{+0.028}_{-0.034}$& 0.132 \\
$n_s$ & $0.951^{+0.015}_{-0.019}$ & 0.983 \\
$A$ & $0.80^{+0.04}_{-0.05}$ & 0.75 \\
  \hline
  \end{tabular}
 \end{center}
\end{table}

\begin{figure}
\begin{center}
\includegraphics[angle=0, width=\hsize]{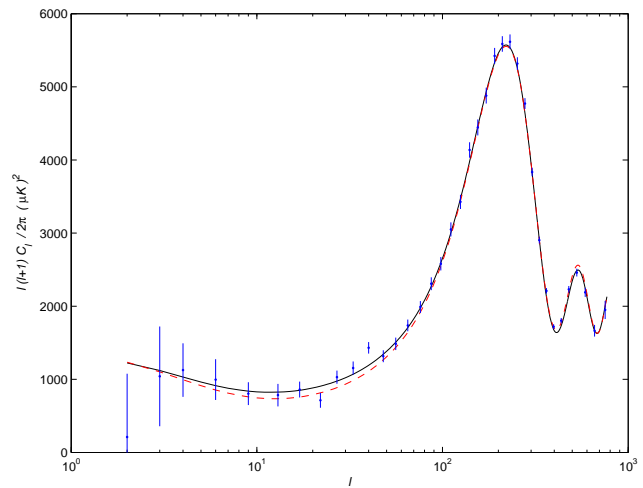}
\caption{The power spectrum for the WMAP best-fit model (solid black
line) and for an illustrative model more consistent with the estimated
variance (red dashed line) are shown. For comparison, the binned power
spectrum measured from the WMAP three-year data (blue points) is also
plotted \citep{spe07}.}
\label{fig:models}
\end{center} 
\end{figure}

In order to find this model, we have carried out the following
analysis. First we have selected a number (around 20) of sets of
values of $\{\tau,n_s,A\}$ which are contained in the most probable 68
per cent of the 3-dimensional likelihood previously constructed for
these parameters. For each of these sets, we have then run CosmoMC
\citep{lew02} in order to find the best-fit to the WMAP data, with
$\tau$, $n_s$ and $A$ fixed for each set, varying the rest of the
cosmological parameters. From all the considered models, we find that
the one that best fits the WMAP power spectrum has the values given in
Table~\ref{tab:models}. It is interesting to note that the value of the
spectral index is higher ($n_s=0.983$) for the new model. As seen in
Figure~\ref{fig:models} the model fits reasonably well the power
spectrum and the main difference can be found at low $\ell$'s where
the new model (red dashed line) goes below the WMAP best-fit (black
solid line), which would tend to give a smaller variance. In
particular, we find that the estimated variance has a theoretical
significance of 93.1 per cent for this new model (to be compared with
99.3 found for the WMAP best-fit) and therefore the CMB variance is
not so significantly low in this case.

Therefore, this seems to indicate that the low value of the CMB
dispersion found in the WMAP data could be made compatible with the
model, while keeping a reasonable fit to the power spectrum, by
varying the cosmological parameters. In addition, we would like to point
out again that we have considered only a restricted set of models and
if the whole set of cosmological parameters were allowed to vary
simultaneously, it is likely that we would find models that would fit
better both the estimated variance and the power spectrum. Nonetheless,
a more exhaustive study where a wider range of models were considered
would be necessary to confirm this result.

\subsection{Gaussianity and isotropy of the CMB}
\label{sec:isotropy}

A deviation of the CMB from Gaussianity and/or isotropy could also
explain the inconsistency found between the estimated CMB variance and
the considered model. Moreover, this possibility is also supported by
the fact that different authors have claimed the presence of
departures from Gaussianity and/or isotropy in the WMAP data (see
\S\ref{sec:intro}).

One possible way to test the Gaussianity and isotropy of the CMB is to
perform a consistency test of the data using equation
(\ref{eq:cmb_disp}), that relates the CMB variance and the power
spectrum. Gaussianity and isotropy have been assumed to estimate both,
the CMB variance and the WMAP power spectrum. Therefore, if
these assumptions are valid, both results should be consistent.

To perform this test, we have used the actual measured power spectrum
of the WMAP data\footnote{The power spectrum measured for the three-year
WMAP data is available at the LAMBDA web site.}, i.e. the power
spectrum estimated for the particular realization of the CMB sky,
instead of the best-fit model. From these $C_{\ell}$'s, we have
calculated the power spectrum that we expect for the WMAP combined map
(using the results of Appendix~\ref{ap:cl_comb_teo}). Finally, we have
calculated the value of the CMB variance from equation
(\ref{eq:cmb_disp}) using this power spectrum of the combined map. In
this way, we find a value for the dispersion of $8.40 \times 10^{-2}$
mK. The error associated to this quantity is expected to be small,
since the main contribution to the variance is due to the low
multipoles, which are almost unaffected by the instrumental noise . This
value should be compared to $(8.16 \pm 0.04) \times 10^{-2}$ mK, the
dispersion that we have previously estimated from the combined
map. The difference between these two estimations indicates the
presence of an inconsistency in the data between the measured power
spectrum and the estimated variance.

To study further this possibility we have also calculated the
theoretical significance of our estimated CMB variance from equation
(\ref{eq:signif_laura}), using again the power spectrum of the
WMAP combined map constructed from the actual measured WMAP power
spectrum (instead of the best-fit model as done in \S\ref{sec:sigma0}). In
this case, we find a value of 96.0 per cent for the
significance. Moreover, if we remove the quadrupole from the data (as
explained in \S\ref{sec:sigma0}), the significance increases to 99.5
per cent\footnote{In \S\ref{sec:sigma0}, when removing the quadrupole,
the significance of the variance decreased (unlike in the current
analysis). This makes sense, since the quadrupole of the data is known
to be significantly lower than that of the best-fit model, and
therefore removing the quadrupole from the analysis would make the
variance more consistent with the best-fit model. However, in the
current analysis, we are comparing the estimated CMB variance with the
actual measured power spectrum of the WMAP data, where the quadrupole
is also low and, therefore, removing the quadrupole, does not
necessarily increase the consistency.}. This points out again towards
an inconsistency between the estimated values of the variance and
those of the power spectrum in the WMAP data (independently of the
best-fit model). In addition, it also shows that the anomaly is
present at multipoles higher than the quadrupole.

Therefore, a deviation from Gaussianity and/or isotropy of the WMAP
data could be the reason for the inconsistency found between the
estimated variance and the considered model. Moreover, taking
into account the results of \S\ref{sec:4}, that show that the
anomalously low variance cannot be explained by known systematics or
foregrounds (and  assuming that the estimated CMB power spectrum
is neither affected by spurious signals), a cosmological origin for
this deviation from Gaussianity and/or isotropy cannot be discarded.

\section{Conclusions}
\label{sec:conclusions}

We have studied the normalised temperature distribution of the CMB
using the WMAP data, finding that the skewness, kurtosis and the
Kolmogorov-Smirnov test are consistent with Gaussianity. However, the
dispersion was significantly lower (at the level of 97.8 per cent)
than expected from CMB simulations using the WMAP best-fit power
spectrum. This result was even more significant when only the northern
ecliptic sky was considered (at 99.4 per cent). In oder to clarify
this point, a more detailed analysis of the CMB dispersion of the WMAP
data has been performed. In particular, we have estimated this
dispersion to be $8.16 \times 10^{-2}$ mK, which is again anomalously
low (98.7 per cent of the Gaussian CMB simulations have a larger
value of the dispersion). If only the northern ecliptic hemisphere is
considered, the significance increases to 99.8 per cent, whereas the
southern ecliptic hemisphere is compatible with the simulations.

We have repeated our analysis using single radiometer and single year
maps, finding that the significance of the low variance is practically
frequency independent and that it cannot be explained by an anomalous
radiometer or by an artifact in a single year data. In addition, the
presence of residual foregrounds or 1/f noise cannot explain the low
value of the CMB dispersion, since they would tend to shift
$\hat{\sigma}_0$ towards higher values.

We have also investigated the effect of varying the cosmological
parameters in the significance of the estimated variance. In
particular, searching within a restricted number of models, we have
fround one case that provides a reasonable fit to the WMAP power
spectrum and for which the signficance of the estimated variance
decreases to 93.0 per cent. This seems to indicate that is possible to
find a model that is consistent with both the estimated variance and
the measured power spectrum of the WMAP data, especially if a wider
range of cosmological models is considered. Nevertheless, a more
exhaustive study is necessary to confirm this result. It is also
interesting to point out that the CMB variance tends to favour higher
values of the spectral index than the full-power spectrum analysis of
the WMAP data.

Finally, we have considered the possibility that a deviation from
Gaussianity and/or isotropy could be the reason for the anomalous
variance. We find that there is an inconsistency between the estimated
CMB variance and the actual measured power spectrum of the WMAP data,
independently of the best-fit model. This indicates that a deviation
from Gaussianity and/or isotropy could be a possible explanation for
the found anomaly. Moreover, taking into account that known
systematics or foregrounds cannot explain the anomalously low CMB
variance, a cosmological origin for this possible deviation cannot be
discarded.

\section*{Acknowledgments}

The authors thank P. Mukherjee for help regarding cosmological
parameter estimation, M. Cruz, A. Curto, D. Herranz,
M. L{\'o}pez-Caniego, and J.L. Sanz for useful discussions and
R. Marco for the computational support. CM thanks the Spanish
Ministerio de Ciencia y Tecnolog\'{\i}a (MCyT) for a predoctoral FPI
fellowship and for two grants to visit The Cavendish Laboratory of the
University of Cambridge (UK). CM, RBB, PV and EMG thank the Cavendish
Laboratory for their hospitality during several research stays and
also acknowledge partial financial support from the Spanish projects
ESP2004-07067-C03-01 and AYA2007-68058-C03-02. Some of the results in
this paper have been derived using the HEALPix (G\'orski et al., 2005)
package.  We acknowledge the use of the Legacy Archive for Microwave
Background Data Analysis (LAMBDA). Support for LAMBDA is provided by
the NASA Office of Space Science. We acknowledge the use of the
software CAMB (http://camb.info/) developed by A. Lewis and
A. Challinor and of CosmoMC developed by A. Lewis and S. Bridle.

\appendix

\section{Theoretical power spectrum of a combined map}
\label{ap:cl_comb_teo}

The derivation of the angular power specturm of the CMB signal
present in the combined Q-V-W map $C_{\ell}^c$ is given in this appendix.
This combined map, $T_c(\vec{x})$, is the
linear combination (see equation \ref{eq:combination})
of the eight cleaned maps ($T_i(\vec{x})$ with $i$ running from 3 to 10)
at different WMAP frequencies (Q1 and Q2 at 41 GHz, V1 and V2 at 61 GHz,
and W1, W2, W3 and W4 at 94 GHz). Let us recall here the
explicit formula:

\begin{equation}
T_{c} ( \vec{x} ) = \sum_{i=3}^{10} {w_i (\vec{x}) T_i(\vec{x})},
\label{eq:comb}
\end{equation}
where $ T_i(\vec{x})$ is the CMB observation made by the $i$
radiometer and the coefficients $w_i (\vec{x})$
represent, at a given postion $\vec{x}$,
the relative weight given to the noise dispersion for each
radiometer, and are normalized to unity (see equation~\ref{eq:noise}).

Each CMB observation made by the $i$ radiometer can be expressed in terms
of the spherical harmonic coefficients of the pure CMB signal $a_{\ell m}$, the
beam ($b_{\ell}^i$) and pixel ($p_{\ell}$) window functions:
\begin{equation}
T_i (\vec{x}) = \sum_{\ell,m}{a_{\ell m} b_{\ell}^i p_{\ell} Y_{\ell m}( \vec{x})}
\label{eq:radiom_map}
\end{equation}

Let us define $\tilde{T_i}(\vec{x}) = w_i (\vec{x}) T_i(\vec{x})$
as the noise-weighted CMB signal observed by the radiometer $i$.
It is straightforward to show that the angular power spectrum of
the combined CMB signal is given by:
\begin{equation}
\label{eq:cl_comb}
C_{\ell}^{c} = \sum_{i,j} {\tilde{C}_{\ell}^{ij}} =
\sum_{i,j} { \frac{1}{2\ell + 1} } \sum_m { \tilde{a}_{\ell m}^i {\tilde{a}_{\ell m}^{j*}} }
\end{equation}
where $\tilde{C}_{\ell}^{ij}$ is the cross-angular power spectrum of
the CMB signal between the $i$ and the $j$ radiometers and
$\tilde{a}_{\ell m}^i$ are the spherical harmonic coeficients of the
noise-weighted CMB signal observed by the radiometer $i$:
\begin{eqnarray}
\label{eq:alm_tilde}
\tilde{a}_{\ell m}^{i} & = &\int {\it{d}\Omega~ \tilde{T}_{i}(\vec{x}) Y_{\ell m}^{*}(\vec{x})}
=  \int {\it{d}\Omega~ w_{i}( \vec{x} ) T_{i}(\vec{x}) Y_{\ell m}^{*} (\vec{x})} = \nonumber \\
&=& \sum_{\ell^{'}m^{'}} a_{\ell^{'} m^{'}} b_{\ell^{'}}^{i} p_{\ell^{'}}
\int{\it{d}\Omega~ w_{i}(\vec{x}) Y_{\ell m}^{*}(\vec{x}) Y_{\ell^{'} m^{'}}(\vec{x})} = \nonumber \\
&=&  \sum_{\ell^{'}m^{'}} a_{\ell^{'} m^{'}} b_{\ell^{'}}^{i} p_{\ell^{'}}
K_{\ell m \ell^{'} m^{'}}^{i}.
\end{eqnarray}
$K_{\ell m \ell^{'} m^{'}}^{i}$ is a matrix introducing coupling among
multipoles due to the noise-weighting.

Including equation~(\ref{eq:alm_tilde}) in equation~(\ref{eq:cl_comb}), the
the angular power spectrum of the combined CMB signal reads:
\begin{eqnarray}
\label{eq:cl_comb2}
C_{\ell}^{c} & = &
\sum_{i,j} { \frac{1}{2\ell + 1} } \sum_m { \tilde{a}_{\ell m}^i {\tilde{a}_{\ell m}^{j*}} } \nonumber \\
&=& \sum_{i,j} { \frac{1}{2\ell + 1} } \sum_m
\left(\sum_{\ell^{'},m^{'}} a_{\ell^{'},m^{'}}  b_{\ell^{'}}^{i} p_{\ell^{'}} K_{\ell m \ell^{'} m^{'}}^{i} \right)
\times \nonumber \\
& &
\left(\sum_{\ell^{''},m^{''}} a_{\ell^{''},m^{''}}  b_{\ell^{''}}^{j} p_{\ell^{''}} K_{\ell m \ell^{''} m^{''}}^{j} \right)^* = \nonumber \\
& = & \sum_{i,j} { \frac{1}{2\ell + 1} } \sum_{\ell^{'}} C_{\ell^{'}} b_{\ell^{'}}^{i} b_{\ell^{'}}^{j}
p_{\ell^{'}}^2 \times \nonumber \\
& &\sum_{m, m^{'}} K_{\ell m \ell^{'} m^{'}}^{i} K_{\ell m \ell^{'} m^{'}}^{j*}.
\end{eqnarray}
where $C_{\ell^{'}}$ is the angular power spectrum of the pure CMB signal and we have applied
$\sum_{\ell^{'},m^{'}} \sum_{\ell^{''},m^{''}} a_{\ell^{'},m^{'}}a_{\ell^{''},m^{''}}^* =
\sum_{\ell^{'},m^{'}} C_{\ell^{'}} \delta_{\ell^{'} \ell^{''}} \delta_{m^{'} m^{''}}$.

The coupling $K_{\ell m \ell^{'} m^{'}}^{i}$ can be expressed in terms of the Wigner $3j$-symbols
and the spherical harmonic coefficients ($w_{\ell m}$) of the noise-weights, as follows:
\begin{eqnarray}
&& K_{\ell m \ell^{'} m^{'}}^{i} = \int{\it{d}\Omega~ w_{i}( \vec{x}
) Y_{\ell m}^{*}( \vec{x} ) Y_{\ell^{'} m^{'}}(
\vec{x} )} = \nonumber \\
&=& \sum_{\ell^{''},m^{''}}{w_{\ell^{''} m^{''}}^{i}
\int{\it{d}\Omega~ Y_{\ell m}^{*}( \vec{x} ) Y_{\ell^{'}
m^{'}}( \vec{x} ) Y_{\ell^{''} m^{''}}( \vec{x} )}} =
\nonumber \nonumber \\
&=& \sum_{\ell^{''},m^{''}} \left\{ w_{\ell^{''} m^{''}}^{i} (-1
)^{m^{'}}
\left[\frac{(2\ell+1)(2\ell^{'}+1)(2\ell^{''}+1)}{4\pi}\right]^{\frac{1}{2}}
\nonumber \right. \nonumber \\
& & \left. \wjjj{\ell}{\ell^{'}}{\ell^{''}}{0}{0}{0}
\wjjj{\ell}{\ell^{'}}{\ell^{''}}{m}{-m^{'}}{m^{''}} \right\}
\label{eq:coupling_matrix}
\end{eqnarray}

One of the orthogonality properties of the Wigner-3j symbols reads
(see e.g. \citealt{hiv02}):
\begin{eqnarray}
\label{eq:ortho3j}
\sum_{m_1,m_2}{\wjjj{\ell_1}{\ell_2}{\ell_3}{m_1}{-m_2}{m_3} \wjjj{\ell_1}{\ell_2}{\ell_4}{m_1}{-m_2}{m_4}}
& = & \nonumber \\
 \frac{1}{2\ell3 + 1} \delta_{\ell_3 \ell_4} \delta_{m_3 m_4} \delta(\ell_1, \ell_2, \ell_3) &&
\end{eqnarray}
where $\delta(l_1, l_2, l_3) = 1$ when the triangular relation $|\ell_1 - \ell_2| \le \ell_3 \le \ell_1 + \ell_2$
is satisfied.

Replacing equation~(\ref{eq:coupling_matrix}) in
equation~(\ref{eq:cl_comb2}) and taking into account
equation~(\ref{eq:ortho3j}), it is straightforward to derive the final
expression for the angular power spectrum of the combined CMB signal:
\begin{eqnarray}
\label{eq:cl_comb3}
C_{\ell}^{c} & = & \sum_{i,j} \sum_{\ell^{'}} C_{\ell^{'}} b_{\ell^{'}}^{i} b_{\ell^{'}}^{j}
p_{\ell^{'}}^2 \frac{2\ell^{'} + 1}{4\pi} \times \nonumber \\
& & \sum_{\ell^{''}} (2\ell^{''} + 1) w_{\ell^{''}}^{ij} \wjjj{\ell}{\ell^{'}}{\ell^{''}}{0}{0}{0}^2,
\end{eqnarray}
where $w_{\ell^{''}}^{ij}$ is the cross-angular power spectrum of the
noise-weights between the radiometers $i$ and $j$.

\begin{figure}
\begin{center}
\includegraphics[angle=0,width=\hsize]{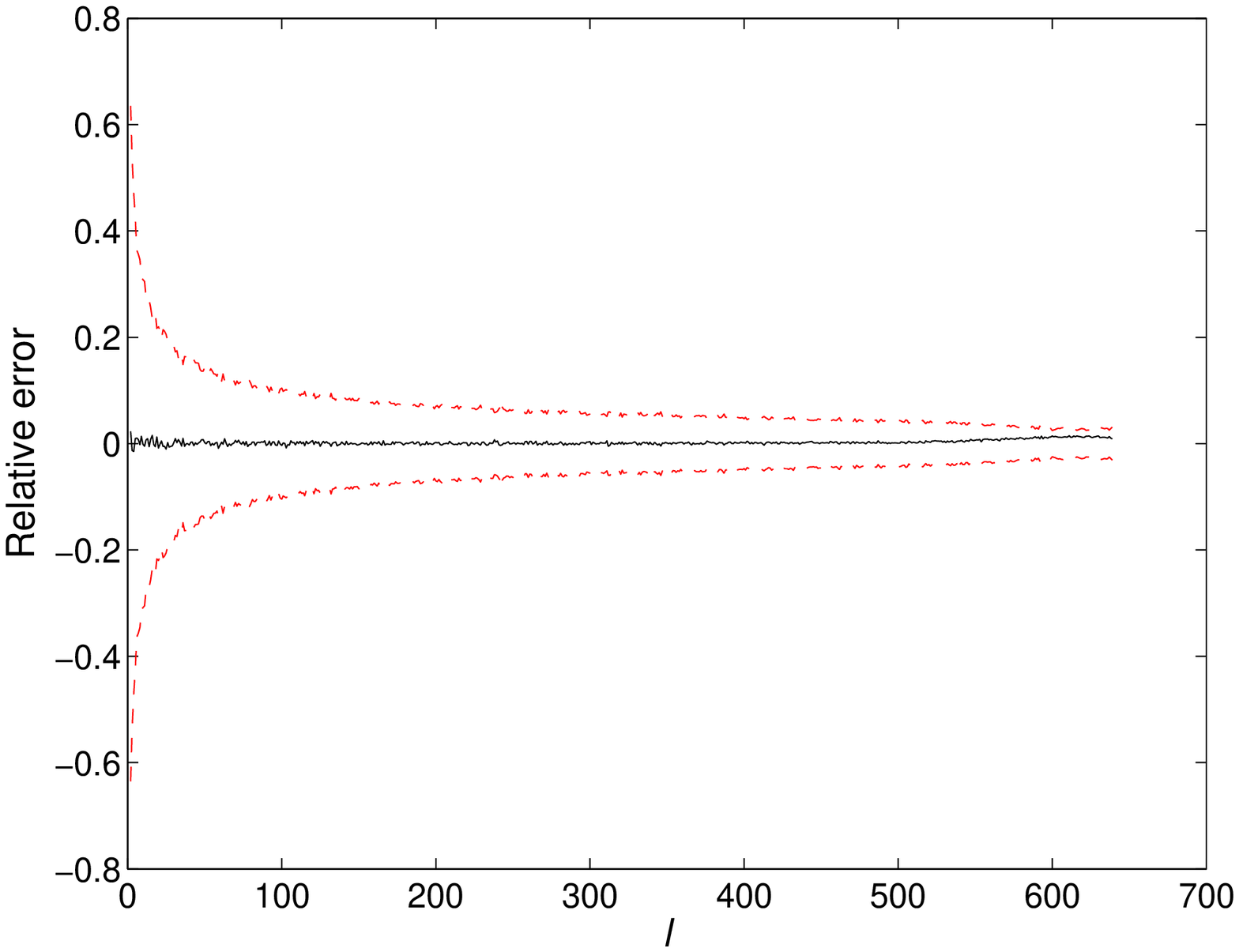}
\caption{Relative error (solid black line) of the theoretical angular
power spectrum of the WMAP combined map, given by equation
(\ref{eq:cl_comb3}), with respect to the one averaged from 1000
simulations. The relative error is obtained as the difference of the
theoretical $C_{\ell}$'s minus the average ones, and this difference
is divided by the average $C_{\ell}$'s. The dashed red lines give the
1$\sigma$ errors (obtained from the simulations) divided by the
averaged angular power spectrum from simulations.}
\label{fig:error}
\end{center} 
\end{figure}

In order to test the performance of this approach,
the theoretical angular power spectrum of the
combined Q-V-W WMAP three-year map (at $n_{side}$=256 resolution)
given by the above equation has been compared with the angular power
spectrum obtained from the average of 1000 Gaussian simulations of
the combined map. In Figure \ref{fig:error}, the relative error
between both spectra is given (solid black line) up to $\ell = 2.5 
n_{side}$. The red dashed lines give the relative 1$\sigma$
error (obtained from the simulations). The agreement between both
angular power spectra in the considered range is very good.

\end{document}